Statistical Tests and Research Assessments:

A comment on Schneider (2012)

Lutz Bornmann* & Loet Leydesdorff**

*Division for Science and Innovation Studies, Administrative Headquarters of the Max Planck Society, Hofgartenstr. 8, 80539 Munich, Germany; bornmann@gv.mpg.de

**Amsterdam School of Communication Research, University of Amsterdam, Kloveniersburgwal 48, NL-1012 CX, Amsterdam, The Netherlands; loet@leydesdorff.net

**Abstract**

In a recent presentation at the 17th International Conference on Science and Technology Indicators (STI) (available at http://2012.sticonference.org/Proceedings/vol2/Schneider_Testing_719.pdf ), Schneider (2012) criticised the proposal of Bornmann, de Moya Anegón, & Leydesdorff (2012) and Leydesdorff & Bornmann (2012) to use statistical tests in order to evaluate research assessments and university rankings. We agree with Schneider's proposal to add statistical power analysis and effect size measures to research evaluations, but disagree that these procedures would replace significance testing. Accordingly, effect size measures (e.g. Cohen's *h*) were added to the Excel sheets that we bring online for testing performance differences between institutions in the Leiden Ranking (at http://www.leydesdorff.net/leiden11) and the SCImago Institutions Ranking (at http://www.leydesdorff.net/scimago11).

*Sir,*

In a recent presentation at the 17th International Conference on Science and Technology Indicators (STI) (available at http://2012.sticonference.org/Proceedings/vol2/Schneider_Testing_719.pdf ), Schneider (2012) criticised our proposal (Bornmann *et al.*, 2012; Leydesdorff & Bornmann, 2012) to use statistical tests in research evaluations and university rankings. We welcome Schneider's suggestion to introduce statistical power analysis and effect size measures, but do not think that these procedures can replace significance testing. With reference to his remarks, we would like to draw attention to the following critical issues.

(*i*) Bornmann *et al.* (2012) and Leydesdorff & Bornmann (2012) proposed to use the *z*-test for determining the statistical significance of results presented in the Leiden Ranking (Waltman *et al.*, in press) and the SCImago Institutions Ranking (SCImago Reseach Group, 2012) when comparing universities or research institutions. Both rankings present the proportion of papers for each institution which belong to the top-10% most highly-cited publications in their subject category and their publication year. This proportion can be tested against the expected value of 10%, since one can expect 10% of thus qualified publications in a sufficiently large random set. Using the *z*-test, one can also compare the proportions of two institutions in terms of the statistical significance of the difference.

Schneider (2012: 720) claims incorrectly that we confused statistical significance with relevance or importance. However, we formulated carefully in the case of the SCImago Institutions Ranking: "The Excellence Indicator allows for testing whether (1) the difference between the institution's percentage and the expected value of 10% or (2) the percentage difference between two institutions is statistically significant" (Bornmann *et al*., 2012, p. 333). With reference to the Leiden Ranking, we formulated: "The statistical significance of performance differences between universities can be tested by using the *z*-test for independent proportions" (Leydesdorff & Bornmann, 2012, p. 781).

(*ii*) Indeed, there is an abundant literature discussing the critique of statistical testing (see Nickerson, 2000, for an overview). However, Schneider (2012) did not mention that this continuous criticism has not resulted in a ban of these tests. In 1999, the American Psychological Association (APA) established a Task Force on Statistical Inference because of this continuing criticism of significance testing. The task was "to elucidate some of the controversial issues surrounding applications of statistics including significance testing and its alternatives; alternative underlying models and data transformation; and newer methods made possible by powerful computers" (Wilkinson, 1999, p. 594). The recommendations of the Task Force were *not* to ban the tests, but to supplement them with statistical procedures such as those proposed by Schneider (2012) (e.g., effect size measures).

The publication manual of APA is authoritative as an editorial standard not only for psychology journals but for more than 1000 other journals in the behavioural, life, and social sciences (Fidler, 2010). Consequently, one continues to use these tests as the standard practice. Statistical significance implies "that the outcome of a study is highly unlikely to have occurred as a result of chance" (Sheskin, 2007, p. 67). Since this does not necessarily suggest that "any difference or effect detected in a set of data is of any practical value" (Sheskin, 2007, p. 67), the statistical tests can be supplemented with power analysis and effect size measures.

In the case of the *z*-test, an appropriate statistic for effect size is Cohen's *h* (Cohen, 1988, pp. 180f.) that measures the difference between the observed ($P_o$) and expected proportions ($P_e$), as follows:

$$h = 2 \times \left| \arcsin \sqrt{P_o} - \arcsin \sqrt{P_e} \right| \quad (1)$$

Since the arcsinus is available as a function in Excel, this value can conveniently be obtained. We extended the Excel sheets for testing the rankings (at http://www.leydesdorff.net/leiden11 and http://www.leydesdorff.net/scimago11, respectively) with these and other measures for the effect size (such as Cramér's *V* and Cohen's *w*; cf. Hadzi-Pavlovic, 2007). Additionally, the user can calculate the standard error and confidence interval for the difference of two proportions.

(*iii*) It is easy to recommend using these procedures, but as Schneider (2012) noted, their interpretation is difficult. For example, Cohen (1988: 216 ff.) suggests replacing Cramér's *V* statistic for contingency tables with *w* values. In the case of *w* values, however, *w*=.1 means a small effect, *w*=.3 a medium effect and *w*=.5 a large effect. Frequently, *w* values in empirical applications lie between these limit values and can then be interpreted as only a "small-to-medium" effect or a "medium-to-large" effect. Then, this effect size measure unfortunately provides little useful information for the interpretation of the results. A similar estimate for *h* values is provided by Cohen (1988: 184f.).

Contrary to Schneider (2012), we recommend that significance tests are not banned from bibliometric research assessment, but that their use can be supplemented, as also suggested by APA (2009: 30): "When applying inferential statistics, take seriously the statistical power considerations associated with the tests of hypotheses. Such considerations relate to the likelihood of correctly rejecting the tested hypotheses, given a particular alpha level, effect size, and sample size."

(*iv*) Schneider (2012) criticises the use of significance tests in research assessment, claiming that they can only be used on random samples. In psychology, experiments are often based on convenience samples and these tests are nevertheless carried out. Researchers at CERN, the European Organization for Nuclear Research, use *p* values to find an answer to the question whether the Higgs Boson really exists (http://for-sci-law-now.blogspot.ch/2012/07/probability-that-higgs-boson-has-been.html) without drawing a random sample first.

Bornmann and Mutz (in press) suggested for evaluative institutional bibliometrics splitting up a population, defined as the whole bibliometric data for an institution, into natural, non-overlapping groups such as different publication years, journals, or authors. Such groups provide clusters in a two-stage sampling design ("cluster sampling") (Levy & Lemeshow, 1999). The clusters should not differ significantly and should exhibit a large heterogeneity in the metrics within the clusters. For example, for an evaluation study, the clusters would consist of three consecutive publication years (e.g. cluster 1: 1990 to 1992, cluster 2: 1993 to 1995 …). Firstly, one single cluster can randomly be selected from the set of clusters. Secondly, all bibliometric data (publications and corresponding metrics) is gathered for the selected cluster. Cluster sampling is particularly efficient if each single cluster represents approximately the whole population. If bibliometric information is available for the whole institution, it is also possible to test whether a selected cluster differs from the population in some important properties.

(*v*) In our opinion, the paper by Schneider (2012) is of further interest to two target groups: (1) *The users of the significance tests* proposed by Bornmann *et al*. (2012) and Leydesdorff & Bornmann (2012): one is now additionally informed about effect sizes in the online sheets (the sheets also provide the various formulas in Excel format). (2) Institutes such as the *Centre for Science and Technology Studies (CWTS)* of Leiden University, which publishes the Leiden Ranking on its web site. Schneider (2012) shows that despite (large) differences in their positions on these rankings, universities do not show significant differences in terms of their performance. For example, he concludes in case of the SCImago Ranking: "In total 678 institutions immediately above or below Stanford University's ranking—ranking positions 61 to 738—have 'trivial' differences in indicator values when compared to Stanford University, according to Cohen's qualitative categories" (Schneider, 2012, p. 726). Bornmann, Mutz, & Daniel (in press) arrived at similar results analysing the Leiden Ranking.

The arguments of Schneider (2012) therefore suggest that CWTS and similar institutes should discontinue the practice of providing users the option to rank universities according to different bibliometric indicators. In the meantime, the SCImago Reseach Group (2012), for example, formulates: "SIR World Report 2012 IS NOT A LEAGUE TABLE. The ranking parameter – the scientific output of institutions – should be understood as a default rank, not our ranking proposal. The only goal of this report is to characterize research outcomes of organizations so as to provide useful scientometric information to institutions, policymakers and research managers so they are able to analyze, evaluate and improve their research results". Furthermore, institutes that produce rankings may in the future wish to follow our example and add suitable statistics to highlight differences such as effect sizes since these can be considered relevant.

Lutz Bornmann and Loet Leydesdorff